\documentclass[aps,prd,superscriptaddress,nofootinbib,eqsecnum,twocolumn]{revtex4-1}

\pdfoutput=1

\usepackage{amsfonts}
\usepackage{amsmath}
\usepackage{amssymb}
\usepackage{graphicx,color}
\usepackage{xcolor}
\usepackage{float}
\usepackage{hyperref}
\usepackage{subfigure}
\usepackage{mathtools,slashed}
\usepackage{blindtext}
\usepackage{longtable}
\usepackage{dcolumn}
\usepackage{bm}
\usepackage{appendix}
\usepackage{multirow}
\usepackage{color}
\usepackage{soul}

\usepackage{lipsum}

\begin{document}

\title{$\alpha$-attractor potentials in loop quantum cosmology}

\author{G. L. L. W. Levy}
\email{guslevy9@hotmail.com}
\affiliation{Centro Brasileiro de Pesquisas F\'{\i}sicas, Rua Doutor Xavier Sigaud 150, Urca, 
22290-180, Rio de Janeiro, RJ, Brazil}

\author{Rudnei O. Ramos} 
\email{rudnei@uerj.br}
\affiliation{Departamento de F\'{\i}sica Te\'orica, Universidade do Estado do 
Rio de Janeiro, 20550-013 Rio de Janeiro, RJ, Brazil}

\begin{abstract}

We perform in this work an analysis of the background dynamics for
$\alpha$-attractor models in the context of loop quantum
cosmology. Particular attention is given  to the determination of the
duration of the inflationary phase that is preceded by the quantum
bounce in these models. {}From an analysis of the general predictions
for these models, it is shown that we can be able to put constraints in the
parameter $\alpha$ of the  potentials and also on the quantum model
itself, especially the Barbero-Immirzi parameter. 
In particular, the constraints on the tensor-to-scalar ratio and spectral tilt
of the cosmological perturbations
limit the $\alpha$ parameter of the potentials to values such that
$\alpha_{n=0} \lesssim 10$, $\alpha_{n=1}
\lesssim 17$ and $\alpha_{n=2} \lesssim 67$,  for the
$\alpha$ attractors T, E, and $n=2$ models, respectively.
Using the constraints on the minimal amount of {\it e}-folds of expansion from the quantum
bounce up to the end of inflation leads to the
upper bounds for the Barbero-Immirzi parameter for the
$\alpha$-attractor models studied in this work:
$\gamma_{n=0} \lesssim 51.2$, $\gamma_{n=1}\lesssim 63.4$ and $\gamma_{n=2} \lesssim 64.2$,
which are obtained when fixing the parameter $\alpha$ in the
potential at the values saturating the upper bounds given above for each model. 

\end{abstract}

\maketitle

\section{Introduction}
\label{intro}

Building a theory of quantum gravity is still a challenge. Among the
proposals, loop quantum gravity (LQG), which is a background
independent and nonperturbative approach  for quantizing general
relativity (for reviews, see,
e.g., Refs.~\cite{Ashtekar:2004eh,Ashtekar:2011ni,Rovelli:2014ssa}) has
been widely investigated in the past 30 years or so. Meanwhile, the
physical implications of LQG make use of its loop quantization
techniques to cosmological models, namely loop quantum cosmology
(LQC), which is the symmetry reduced version of
LQG~\cite{Ashtekar:2003hd,Bojowald:2005epg,Ashtekar:2006rx,Ashtekar:2009mm,Ashtekar:2011ni,Barrau:2013ula,Agullo:2016tjh}. In
LQC, the quantum effects at the Planck scale are able to produce a
bounce that results as a consequence of the repulsive quantum
geometrical effects and, hence, effectively resolving the singularity
issue of classical general relativity. 

Making predictions concerning the inflationary phase that can be
preceded by a quantum bounce has attracted quite some interest
recently. Interestingly, it has been shown that in LQC that inflation
can occur quite naturally, and it is in general a strong attractor when a
scalar field is the main ingredient of the energy density. This
characteristic of LQC has been confirmed and studied in detail in
many recent
works~\cite{Zhu:2017jew,Shahalam:2017wba,Li:2018fco,Sharma:2018vnv,Li:2019ipm,Shahalam:2019mpw}. These
works have shown that in LQC models with a kinetic energy dominated
bounce lead to an almost inevitably inflationary phase following the
bounce phase.

The discussion of the inflationary phase after the quantum bounce in
LQC has been mostly studied following two lines of thoughts on how and
when the initial conditions should be taken.  One line of thought
assumes that the initial conditions can be appropriately taken at the
bounce~\cite{Ashtekar:2011rm,Zhu:2017jew,Shahalam:2017wba,Li:2018fco,Sharma:2018vnv,Graef:2018ulg,Li:2019ipm,Shahalam:2019mpw}.
Another school assumes that the appropriate moment to take the initial
conditions would be deep inside the contracting phase before the
bounce~\cite{Linsefors:2013cd,Linsefors:2014tna,Bolliet:2017czc,Martineau:2017sti,Barboza:2020jux}.
Both lines of thought lead to the conclusion that inflation is in
general a strong attractor; however, in the latter case, when taking
initial conditions deep in the contracting phase, it has been further
demonstrated that not only inflation is highly probable, but that  the
duration of inflation itself can be predicted using simple analytical
methods as has been shown in Ref.~\cite{Barboza:2022hng}. Having a way
of making predictions concerning the inflationary phase is quite important
when comparing and contrasting different inflationary models with the
observations. 

In the present work, we make use of the method developed in
Ref.~\cite{Barboza:2022hng} and apply it to the study of the dynamics
for the $\alpha$-attractor type of
potentials~\cite{Kallosh:2013hoa,Kallosh:2013yoa,Kallosh:2013tua}. We
focus in particular on the bouncing dynamics and the subsequent
transition (preinflation phase) and inflationary phases following the
quantum bounce in LQC. {}For different values of parameters in these
potentials, we verify whether they can produce not only a sufficient number of
{\it e}-folds such as to solve the usual big bang problems but also to be
compatible with the observational predictions for these types of
models, e.g., the tensor-to-scalar ratio and spectral tilt of the
scalar perturbations.  We note that the preinflationary dynamics in
LQC for this class of potentials has also been studied previously  in
Refs.~\cite{Shahalam:2018rby,Shahalam:2019mpw} by adopting the
school of thought of taking the initial conditions at the instant of the
bounce, as proposed e.g., in Ref.~\cite{Ashtekar:2011rm}. In this
approach, the authors of those references have then checked which
initial conditions at the bounce would be able to lead to a sufficient
duration of inflation in different $\alpha$-attractor models. 
Here, however, we follow the
second line of thought, which argues that the appropriate instant for
taking the initial conditions should be  in the
contracting phase and well before the bounce, as initially proposed in
Ref.~\cite{Linsefors:2013cd}. As already commented above, this has the
additional advantage of making it possible to make precise predictions 
for what should be
the actual duration of inflation and avoids the arbitrariness of the
former line of thought, of which initial condition should one actually take at
the bounce instant. 

It is important to comment that in this work, like in the previous 
literature~\cite{Linsefors:2013cd,Linsefors:2014tna,Bolliet:2017czc,Martineau:2017sti,Barboza:2020jux,Barboza:2022hng},
we are assuming the vanillalike old LQC description where the deep contracting phase
behaves like contracting classical general relativity. But we have to keep in
mind that there are other examples of LQC models where the prebounce physics
can be much more complicated, like
involving a collapsing universe with a large cosmological constant
(of Planck order)~\cite{Dapor:2017rwv,Agullo:2018wbf}, or that also 
involves spacetimes that are emergent~\cite{Alesci:2016xqa,Olmedo:2018ohq}, i.e., with a
transition from Minkowski to LQC-FRW. All these models can be considered as motivated
from the full theory and can be as well natural extensions of LQC.
The results derived in this paper cannot be extended to those other formulations of
LQC where the most adequate point of setting the initial conditions might be at the bounce
instant and where predictions like the ones we have obtained cannot the derived.

The interest in considering the $\alpha$-attractor type of
potentials is because they are a well-motivated class of inflationary
potentials which are able to lead to universal predictions for
large-scale observables that are largely independent of the details of
the inflationary potential. {}Furthermore, they lead to predictions
that lie close to the center of current observational bounds on the
primordial power spectra~\cite{Planck:2018jri,Planck:2019nip}.  Let us
recall that the Starobinsky potential~\cite{Starobinsky:1980te}, which
is shown to fit the current observations quite well,  is a particular
case of an $\alpha$-attractor type of potential. 

We have organized this paper as follows. In Sec.~\ref{sec2} we briefly
review the  $\alpha$-attractor, in particular the T, E, and $n=2$ models
which are considered in this work.  In Sec.~\ref{sec3} we
present the structure of LQC for developing the background dynamics
and summarize the methods developed in Ref.~\cite{Barboza:2022hng},
which are used in the present study.  Our main results are presented
in Sec.~\ref{sec4}, where we study the background dynamics of the 
T, E, and $n=2$ models, which includes the bouncing, preinflation, and inflation
phases and present the predictions that we obtain for these models. We
also contrast our results with the previous ones obtained when
considering initial conditions taken at the bounce instant. We use these results
to constrain the $\alpha$ parameter of the potentials and also the Barbero-Immirzi
parameter when considering it as a free parameter in LQC.
{}Finally, in Sec.~\ref{conclusions}, we present our conclusions.
Relevant technical details used in our analysis are also given in the
Appendix.

\section{The $\alpha$-attractor potentials}
\label{sec2}

\begin{center}
\begin{figure}[!htpb]
\includegraphics[width=8.2cm]{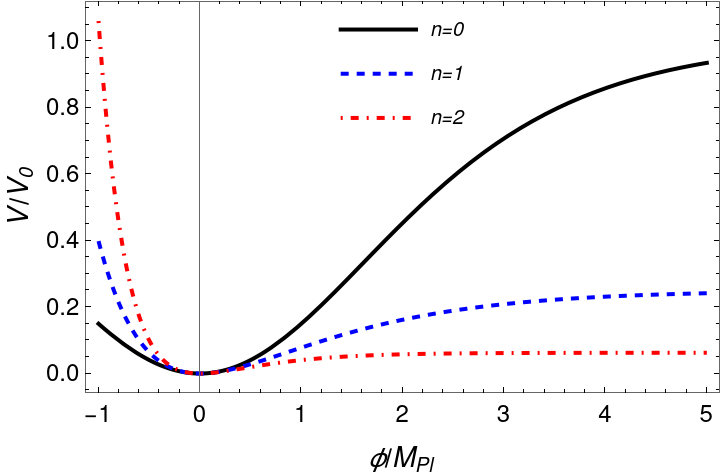}
 \caption{The $\alpha$-attractor potential cases considered in this
   work. The parameter $\alpha$ was set to the value $\alpha=1$.}
 \label{fig:potential}
 \end{figure}
\end{center}

In this paper, we consider the following class of $\alpha$-attractor
potentials given
by~\cite{Linder:2015qxa,Shahalam:2016juu,Shahalam:2018rby} 
\begin{equation}
\label{alphattractor}
V(\phi) = V_0 \frac{\left[\tanh{\left(\frac{\phi}{\sqrt{6\alpha}\,
        M_{\rm
          Pl}}\right)}\right]^2}{\left[1+\tanh{\left(\frac{\phi}{\sqrt{6\alpha}
        \, M_{\rm Pl}}\right)}\right]^{2n}},    
\end{equation}
where $M_{\rm Pl} =m_{\rm Pl}/\sqrt{8\pi}$ is the reduced Planck mass
and $m_{\rm Pl}=1.22 \times 10^{19}\,$GeV is the Planck mass, and $\alpha$
is a dimensionless positive constant. The value of the power $n$
parametrizes different classes of $\alpha$-attractor potentials. We
will work with the cases where $n=0, \, 1,\, 2$. {}For instance, for
$n=0$, in the literature, this potential is called the
T model~\cite{Kaiser:2013sna,Kallosh:2013yoa,Kallosh:2014rga}.  The
case with $n=1$ is know as the E model and it is a generalization of
the Starobinsky model~\cite{Starobinsky:1980te}, which is obtained
when $\alpha = 1$.  In Eq.~(\ref{alphattractor}), the value of $V_0$
is the normalization of the potential, which is fixed by the amplitude
of the CMB scalar power spectrum for given values of $n$ and $\alpha$
(see the Appendix for details).  These types of potentials are well
motivated for describing dark energy models to explain the late time
cosmic acceleration~\cite{Dimopoulos:2017zvq}, and they are also forms of
potentials representing a limiting case of more general modified 
gravity theories, like the Starobinsky form. {}For large values of $\alpha$ 
the models resemble monomial potentials~\cite{Kallosh:2013yoa}, and if 
$\phi \ll M_{\rm Pl}$ the potentials approach a quadratic form, while for $\phi \gg
M_{\rm Pl}$ they approach a constant (flatten) form. The three cases
we will consider, $n=0,\,1$, and $2$ are shown in
{}Fig.~\ref{fig:potential} for illustration.

These potentials also represent robust inflationary models when regarding
their predictions. When compared with the observational constraints,
like with the  spectral tilt $n_s$ and the tensor-to-scalar ratio $r$, they are
able to fit the data well. {}For instance, when $\alpha \ll 1$ and a
large number of {\it e}-folds of inflation they lead
to~\cite{Carrasco:2015pla} (see also the Appendix for details)
\begin{equation}
n_s \simeq 1- \frac{2}{N_*},
\label{ns}
\end{equation}
and
\begin{equation}
r \simeq \alpha \frac{12}{N_*^2},
\label{r}
\end{equation}
where $N_*$ is the number of {\it e}-folds corresponding to the scales
crossing the Hubble radius relevant to CMB, typically corresponding to 
$N_* = 50 - 60$ {\it e}-folds before the end of inflation, depending on the reheating
history~\cite{Liddle:2003as}.  In the absence of running, the Planck
data measure the spectral index to be~\cite{Planck:2018jri}
\begin{equation}
n_s = 0.9649\pm 0.0042,
\label{valuens}
\end{equation}
while the recent data analysis of the BICEP, Keck Array combined with that
from Planck data places the upper bound on the tensor-to-scalar
ratio~\cite{BICEP:2021xfz},
\begin{equation}
r < 0.036, \;\;\;\;{\rm at}\;\; 95\%CL.
\label{valuer}
\end{equation}

In the Appendix, we show the behavior of both $r$ and $n_s$ for
the three types of $\alpha$ attractors studied here and which are
valid for any value of $\alpha$.  It is found that the most
constraining condition comes from the tensor-to-scalar ratio, giving
the upper bounds on $\alpha$ obtained, e.g., at the lowest value of
$N_*=50$,
\begin{eqnarray}
\alpha \lesssim \left\{
\begin{array}{ll}
10, & {\rm for}\; n=0,\\ 17, & {\rm for}\; n=1,\\ 67, & {\rm for}\;
n=2.
\end{array}
\right.
\label{alphabound}
\end{eqnarray}

\section{Background dynamics in LQC}
\label{sec3}

In LQC, the {}Friedmann equation is modified by the quantum effects and
given by~\cite{Ashtekar:2009mm,Ashtekar:2011rm}
\begin{equation}\label{friedmann}
H^{2} = \frac{8 \pi}{3 m^{2}_{\rm Pl}} \rho \; \left(1 -
\frac{\rho}{\rho_{cr}}\right),
\end{equation}
where $\rho$ is the total energy density and $\rho_{cr}$ is the
critical energy density at which the bounce happens. {}For $\rho \ll
\rho_{cr}$ we recover general relativity as expected. The critical
energy density is given by
\begin{equation}
\rho_{cr} =  \frac{\sqrt{3}m_{\rm Pl}^4}{32 \pi^2 \gamma^3}, 
\label{rhoc}
\end{equation} 
and with $\gamma$ being the Barbero-Immirzi parameter. It is common in
the literature of LQC to assume  the Barbero-Immirzi parameter as
given by the value $\gamma \simeq 0.2375$, which is motivated by black
hole entropy calculations~\cite{Meissner:2004ju}. However, many
authors prefer to consider $\gamma$ to be a free parameter in quantum
gravity theories (see,
e.g., Refs.~\cite{Engle:2010kt,Bianchi:2012ui,Wong:2017pgl}). In our
analysis, to be performed in the next sections, we will consider both
points of view. In particular, when taking $\gamma$ as a free
parameter, we will explore how the type of potentials we are
considering in LQC can be made consistent with the observations. This
will allow us to put an upper bound in $\gamma$.

In this present paper, we work with the dynamics of one scalar field
$\phi$, the inflaton, with the potential as given by
Eq.~(\ref{alphattractor}) in the three cases mentioned previously, the
T model ($n=0$) the E model ($n=1$), and the case with $n=2$.  In the
{}Friedmann-Lema\^itre-Robertson-Walker metric, the
background evolution for the inflaton is given by 
\begin{equation}
 \ddot{\phi} + 3H \dot{\phi} + V_{,\phi} = 0,
\label{eom}
\end{equation} 
where $V_{,\phi} \equiv dV(\phi)/d\phi$ is the derivative of the
inflaton's potential. 

As stated in the Introduction~\ref{intro}, we follow the background dynamics
starting from the contracting phase, well before the bounce, where the
initial conditions are set, and follow the dynamics of the inflaton
through the bounce, along the postbounce expanding preinflationary
phase, the beginning and end of inflation.   We follow closely the
derivation considered in Ref.~\cite{Barboza:2022hng} for each one of
these dynamical phases and for which the detailed analysis was
provided. As in the previous references analyzing the  dynamics of
inflation after the bounce, we will always be assuming that the bounce
is dominated by the kinetic energy of the inflaton (which is in fact a 
natural condition when the initial conditions are taken deep in the
contracting phase as shown in Ref.~\cite{Barboza:2022hng}). Since the kinetic
energy evolves like a stiff fluid, $\dot\phi^2 \propto 1/a^6$,
it will generically dominate over the potential energy density 
at the bounce when starting with
initial conditions for the inflaton deep in the contracting phase.
Below we will summarize the main equations for each one of the 
phases that will be important for our study.  

\subsection{Setting the initial conditions in the contracting phase}
\label{preanalysis}

We set the initial conditions in the classical contracting phase at
some instant $t$ well before the bounce time $t_B$ and for which the
quantum effects are still negligible. In this case, the Hubble
parameter can be expressed as
\begin{equation}
H \simeq \frac{1+\beta}{3 (t-t_B)},
\label{Halpha}
\end{equation}
where $\beta$ defines here the ratio between potential and kinetic
energy densities for the scalar field, $\beta = V/(\dot \phi^2/2)$.
\footnote{
Please note that in Ref.~\cite{Barboza:2022hng}
the notation $\alpha$ was used for the ratio between potential and kinetic
energy densities for the scalar field. We have changed the notation 
to avoid confusion with the $\alpha$ parameter in the potentials considered
here.}
On the other hand, as shown in Ref.~\cite{Barboza:2022hng}, 
the time interval 
between some instant $t_\beta$ in the contracting phase for a given value of $\beta$
and the bounce instant $t_B$
can also be expressed as
\begin{equation}
t_\beta - t_B = - \frac{1+\bar \beta}{3}  \sqrt{ \frac{3 m_{\rm Pl}^2
    \bar \beta}{8 \pi (1+\bar \beta) V(\phi_\beta)}},
\label{tbeta}
\end{equation}
where $\phi_\beta \equiv \phi(t_\beta)$ and $\bar{\beta}$ is taken as
the ``average" value for $\beta$, and we approximate it as a constant
within the range $(0, 1)$ (see Ref.~\cite{Barboza:2022hng} for
details). Here, the choice of $\beta$ parametrizes how far in the
past we set the initial conditions for the inflaton field. Once the
potential $V(\phi)$ is specified, $\phi_\beta$ is obtained by the
solution of
\begin{equation}
\frac{V(\phi_\beta)}{V'(\phi_\beta)} = \frac{\sqrt{1+\bar \beta}}{4
  \sqrt{3 \pi}}   m_{\rm Pl}.
\label{Vratio}
\end{equation}
As shown in Ref.~\cite{Barboza:2022hng} (for a similar earlier
prescription, see also Ref.~ \cite{Ashtekar:2011rm}), there is one
value of $\bar \beta$ that can be fixed once and for all for all
potentials, given by $\bar \beta =1/3$ and which is the value
we will be using throughout our analysis.
This value of $\bar \beta =1/3$ was determined in 
Ref.~\cite{Barboza:2022hng} by comparing the numerical results for
the number of {\it e}-folds of inflation for different potentials obtained 
by a statistical analysis performed in Ref.~\cite{Barboza:2020jux},
which considered 
a large number of random initial conditions deep in the contracting phase 
and each one of those initial conditions evolved up to the end of
inflation. This allowed for a probability distribution function for each
potential to be obtained, from which statistical predictions for
the number of {\it e}-folds were derived. As shown in Ref.~\cite{Barboza:2022hng},
the comparison of those numerical results with the analytical ones
obtained using $\bar \beta =1/3$ agree quite well, with overall differences
which are less than 5$\%$.

\subsection{The bounce phase}

The solution $\phi_\beta$ can be connected with the valid one around
the bounce phase. As at the bounce phase we generically expect
$\dot{\phi}^2/2\gg V$, thus leading to
\begin{equation}
\ddot{\phi} + 3H \dot{\phi}  \approx 0,
\label{eom2}    
\end{equation}
and we can solve Eq.~(\ref{eom2}) when using
(\ref{friedmann}), with the initial condition $\phi(t_B)=\phi_B$, to
obtain~\cite{Zhu:2017jew}
\begin{equation}\label{mov_evol}
\phi(t) = \phi_{\rm B} \pm \frac{m_{\rm Pl}}{2\sqrt{3
    \pi}}\,\operatorname{arcsinh}{\left[\sqrt{\frac{24 \pi \rho_{\rm
          cr}}{m_{\rm Pl}^4}}\,\frac{(t-t_{\rm B})}{t_{\rm
        Pl}}\right]}.
\end{equation}
As shown in Ref.~\cite{Barboza:2022hng}, the solution
given by Eq.~(\ref{mov_evol}) holds well even deep in the contracting
phase. This allows one to determine $\phi_B$ once $\phi_\beta$ is
obtained.

\subsection{The postbounce preinflationary phase}

After the bounce, in the expanding phase the kinetic energy of the
inflaton dilutes faster than its potential energy. We denote this
phase, that lasts from the bounce instant $t_B$ up to the transition point 
$t_{tr}$, where the
potential energy equates to the kinetic energy (i.e., $\dot \phi^2(t_{tr})/2 =
V(\phi(t_{tr}))$, or $w=0$),  as the postbounce preinflationary phase.  
The inflaton's amplitude in the transition time is 
\begin{equation}\label{mov_transition}
\phi(t_{tr}) = \phi_{\rm B} + \frac{m_{\rm Pl}}{2\sqrt{3
    \pi}}\,\operatorname{arcsinh}{\left[\sqrt{\frac{24 \pi \rho_{\rm
          cr}}{m_{\rm Pl}^4}}\,\frac{(t_{tr}-t_{\rm B})}{t_{\rm
        Pl}}\right]},
\end{equation}
and the time at this transition point $t_{tr}$ is
determined by solving
\begin{equation}
\dot{\phi}(t_{tr}) = \sqrt{2 V(\phi(t_{tr}))},
\label{dotphiV}
\end{equation}
where we choose the convention of positive sign in both 
Eqs.~\eqref{mov_evol} and  \eqref{dotphiV}  as explained in 
Ref.~\cite{Barboza:2022hng}. We note that explicit analytical 
expressions for both $t_{tr}-t_B$ and
$\phi(t_{tr})$ can be obtained from these equations by approximating
them by considering that $t_{tr}-t_B \gg t_{\rm Pl}$ as shown in
Ref.~\cite{Zhu:2017jew}. However, these expressions are in general too
complicated, and since Eq.~(\ref{dotphiV}) is valid for any potential,
it is simpler just to  directly solve it numerically, as we do here. 

\subsection{The inflationary phase}

Soon after the transition phase, inflation starts.  The instant of the
start of the accelerating inflationary regime, $t_i$, is given when
$w=-1/3$, i.e., when $\dot\phi^2_i = V(\phi_i)$.  The time interval between
the transition phase and the beginning of inflation has been shown to
be very short~\cite{Zhu:2017jew}, lasting much less than one {\it e}-fold and
can be neglected. This allows us to obtain $\phi_i \equiv \phi(t_i)$
as
\begin{equation}\label{phi_initial}
    \phi_{i} \simeq \phi_{tr} + \dot{\phi}_{tr}
\,t_{tr}\,\ln{\frac{t_{i}}{t_{tr}}}.
\end{equation}
{}Finally, from the   slow-roll coefficient $\epsilon_V$,
\begin{equation}
\epsilon_V =\frac{m_{\rm Pl}^2}{16 \pi }
\left(\frac{V'}{V}\right)^{2},
\label{epsilon}
\end{equation}
we determine the inflation amplitude at the end of inflation,
$\phi_{\rm end}$, by setting  $\epsilon_V=1$.  {}For the
$\alpha$-attractor class of potentials considered here,
Eq.~(\ref{alphattractor}), we obtain that
\begin{widetext}
\begin{eqnarray}
\phi_{\rm end}= \frac{1}{2} \sqrt{\frac{3 \alpha}{\pi}} \, m_{\text{Pl}}  \,
\text{arccoth}\left[\frac{1}{2}  \left(n+ \sqrt{3\alpha
  }+\sqrt{(-2+n)^2+2 n\sqrt{3 \alpha}  +3 \alpha}\right)\right].
\label{phiend}
\end{eqnarray}
\end{widetext}

The  total number of {\it e}-folds of inflation is then given by
\begin{equation}
   N_{\rm infl} \approx \frac{8\pi}{m_{\rm Pl}^2}  \int_{\phi_{\rm
       end}}^{\phi_i} \frac{V}{V'} d\phi.
\label{Nefolds}
\end{equation}

\section{Results}
\label{sec4}

Let us now present our results obtained from the analysis of the
$\alpha$-attractor models considered here. As a preliminary analysis,
we consider the case where the initial conditions are set at the
bounce instant $t_B$, according to the philosophy  adopted in
Refs.~\cite{Shahalam:2018rby,Shahalam:2019mpw} in the study using
these types of inflaton potential. This will allow us to confront how
well the analytical results produced for these potentials perform when
compared with the numerical ones, obtained by a direct numerical
solution of the background evolution equation for the inflaton,
Eq.~(\ref{eom}), with the modified Friedmann equation
(\ref{friedmann}) in LQC.   Note that in this case, we set arbitrary
values for the inflaton amplitude $\phi_B$ at the bounce, but still
subjected to the condition that at the bounce the kinetic energy would
dominate over the potential energy of the inflaton, $\dot \phi_B^2/2
\gg V(\phi_B)$.  After this preliminary study, we will follow the 
philosophy that the initial
conditions should be taken deep in the contracting phase and follow
the dynamics from this point on up to the end of inflation.  In this
case, we follow the methodology presented in
Ref.~\cite{Barboza:2022hng} and summarized in the previous section. In
these first two analyses, we work with a Barbero-Immirzi parameter
that is fixed at the value $\gamma =0.2375$ as motivated by black hole
thermodynamic studies. {}Finally, we will let $\gamma$  vary and
determine how the dynamics of the  $\alpha$-attractor models can lead
to constraints on its value when confronted with the observations.

\subsection{Initial conditions set at the bounce}

\begin{table*}[!htpb]
\centering
\caption{Comparison between the numeric and analytic solution for T ($n=0$), E ($n=1$), and
$n=2$ $\alpha$-attractor models.}
\begin{tabular}{ccccccccccc}
\hline
Model  &  &  & $\phi_{\rm B}/m_{\rm Pl}$ &  & $\phi_{\rm tr}/m_{\rm Pl}$ &  & $\phi_{\rm i}/m_{\rm Pl}$ &  & $N_{\rm infl}$ &  \\ \hline \hline
$n=0$ (numeric)  &  &  & $0.10$                   &  & $2.35$                    &  & $2.39$                   &  & $178.3\;(144.3)$      &  \\
$n=0$ (analytic) &  &  & $0.10$                   &  & $2.38$                    &  & $2.44$                   &  & $158.5$       &  \\ 
$n=1$ (numeric)  &  &  & $0.10$ &  & $2.37$ &  & $2.41$ &   & $355.1\; (287.0)$ &  \\
$n=1$ (analytic) &  &  & $0.10$ &  & $2.40$ &  & $2.46$ &   & $316.2$ &  \\ 
$n=2$ (numeric)  &  &  & $0.10$ &  & $2.34$ &  & $2.38$ &  & $210.4\; (172.1)$ &  \\
$n=2$ (analytic) &  &  & $0.10$ &  & $2.38$ &  & $2.43$ &   & $162.7$ & \\ \hline \hline
\label{tab1}
\end{tabular}
\end{table*}

\begin{table*}[!htpb]
\centering
\caption{The predicted results for $\phi_B$, $\phi_{\rm tr}$, $\phi_{\rm i}/m_{\rm Pl}$, and $N_{\rm infl}$  
for the three forms of the $\alpha$-attractor models. }
\begin{tabular}{cccccccccc}
\hline \hline
Models &  &  & $\phi_{\rm B}/m_{\rm Pl}$ &  & $\phi_{\rm tr}/m_{\rm Pl}$ &  & $\phi_{\rm i}/m_{\rm Pl}$ &  & $N_{\rm infl}$ \\ \hline 
\begin{tabular}[c]{@{}c@{}}T-model ($\alpha=5$)\end{tabular}   &  &  & $2.69$                   &  & $4.97$                    &  & $5.02$                   &  & $1.84\times 10^4$      \\
\begin{tabular}[c]{@{}c@{}}E-model ($\alpha=5$)\end{tabular}   &  &  & $2.62$                   &  & $4.92$                    &  & $4.97$                   &  & $3.36 \times 10^4$      \\
\begin{tabular}[c]{@{}c@{}}$n=2$ model  ($\alpha=30$)\end{tabular} &  &  & $2.62$                   &  & $4.89$                    &  & $4.95$                   &  & $8.87\times 10^3$      \\ \hline \hline
\label{tab2}
\end{tabular}
\end{table*}

By setting the initial conditions at the bounce, we fix $\phi_B$ and
then using Eqs.~(\ref{mov_evol}), (\ref{dotphiV}), (\ref{phi_initial}),
and (\ref{phiend}), we evaluate the inflaton's amplitude at the
transition point, at the beginning and at the end of inflation. We
denote these results as being the analytical ones. The assumed value
for $\phi_B$ is such that the number of {\it e}-folds produced are not too
large and, thus, we can better control the numerical solution as far as
precision and time of evaluation are concerned. Then, the numerical
results are obtained by using the same value for $\phi_B$, but
numerically evolving Eq.~(\ref{eom}) with the {}Friedmann equation
(\ref{friedmann}) and obtained in Ref.~\cite{Shahalam:2018rby}. We
also choose the initial conditions such that the evolution is always
at the flatter region of the $\alpha$-attractor models (see
{}Fig.~\ref{fig:potential}), i.e., $\phi_B>0$ and $\dot \phi_B >0$.
The results for each of the corresponding phases and the number of
{\it e}-folds for the inflationary phase for the T-model ($n=0$), E-model
($n=1$), and for the $n=2$ $\alpha$-attractor models are shown in
Table~\ref{tab1}. To obtain these results, we have considered the
parameter value $\alpha$ for each model as set to $\alpha_{n=0}=5$,
$\alpha_{n=1}=5$, and $\alpha_{n=2}=30$,  whose values are consistent
with the ones given by the upper bounds given in
Eq.~(\ref{alphabound}).  The normalization $V_0$ for each potential is
computed according to the Appendix and always at the $N_*=50$ value
for definiteness.

In Table~\ref{tab1}, the result for $N_{\rm infl}$ 
in parentheses for the numerical results indicates the inflationary number of {\it e}-folds computed
according to the slow-roll approximation, Eq.~(\ref{Nefolds}), whose formula we also used
when estimating the analytical results. The number of {\it e}-folds without the parentheses is obtained when we follow
the exact slow-roll coefficient $\epsilon_H = - \dot H/H^2$ when it first becomes equal to $1$ in the
expanding phase after the bounce (e.g., $w_\phi = -1/3$) and when it gets equal to $1$ again later at the
end of inflation.
As noticed by the results shown in Table~\ref{tab1}, the largest differences
come from the obtained number of {\it e}-folds in each method. The slow-roll formula,  Eq.~(\ref{Nefolds}), 
produces results with a difference between the analytic and numerical results that is of order of $10\%$.

\subsection{Initial conditions set in the far past in the contracting phase}

We now consider the line of thought that the appropriate moment to
take the initial conditions should be deep inside the contracting
phase before the
bounce~\cite{Linsefors:2013cd,Linsefors:2014tna,Bolliet:2017czc,Martineau:2017sti,Barboza:2020jux}. 
We follow the evolution of the inflaton field from deep inside
the contracting phase, before the bounce, until up to the end of inflation, 
according to the method explained in
Ref.~\cite{Barboza:2022hng} and summarized in the previous section,
Sec~\ref{sec3}. As shown in Ref.~\cite{Barboza:2022hng}, provided the
initial conditions are set sufficiently  far back in the contracting
phase, the bounce will always be dominated by the kinetic energy of
the inflaton, which always grows much faster than the  energy density
in the potential of the inflaton. This then allows one to uniquely compute
and determine the evolution of the inflaton up to the end of
inflation. In particular, the total number of {\it e}-folds of inflation
$N_{\rm infl}$ becomes a predicted quantity for a given potential.
The obtained results for the $\alpha$-attractor potentials considered
in this work are shown in Table~\ref{tab2}.  Here, we once again fixed
the Barbero-Immirzi parameter at the value $\gamma =0.2375$ and
considered the values for the parameter $\alpha$ as given in the
previous analysis above,  $\alpha_{n=0}=5$, $\alpha_{n=1}=5$, and
$\alpha_{n=2}=30$. The cases of varying $\alpha$ and $\gamma$ will
also be considered below. 

\begin{center}
\begin{figure}[!htpb]
\includegraphics[width=8.2cm]{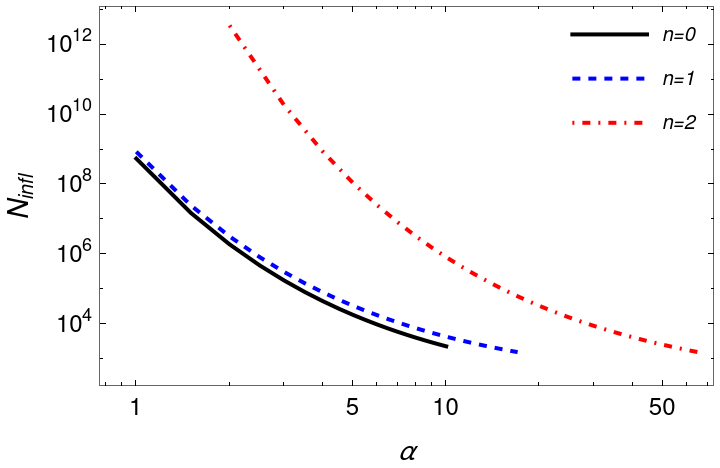} 
 \caption{Number of inflationary {\it e}-folds $N_{\rm infl}$ for the
   $\alpha$-attractor potentials as a function of the $\alpha$
   parameter. The Barbero-Immirzi parameter is kept at the value
   $\gamma =0.2375$. The upper values of $\alpha$ considered for each
   model follow from Eq.~(\ref{alphabound}).}
 \label{fig2}
 \end{figure}
\end{center}

In {}Fig.~\ref{fig2}, we still consider the Barbero-Immirzi parameter
fixed at $\gamma =0.2375$, but study how the results for the number of
{\it e}-folds of inflation $N_{\rm infl}$ changes by varying the parameter
$\alpha$ for each model. The maximum value for $\alpha$ for each model
is taken as given by the upper bounds given in Eq.~(\ref{alphabound})
such that the $\alpha$-attractor models are consistent with the
observations.

We note from the results shown in {}Fig.~\ref{fig2} that the smaller the parameter $\alpha$ is, the larger the number of {\it e}-folds
predicted for each model is. This is consistent with the
fact that the smaller is $\alpha$, the region of the potentials 
where inflation begins gets flatter, thus generically leading to 
a larger number of {\it e}-folds of inflation.

\subsection{Varying the Barbero-Immirzi parameter}

\begin{center}
\begin{figure}[!htpb]
\includegraphics[width=8.2cm]{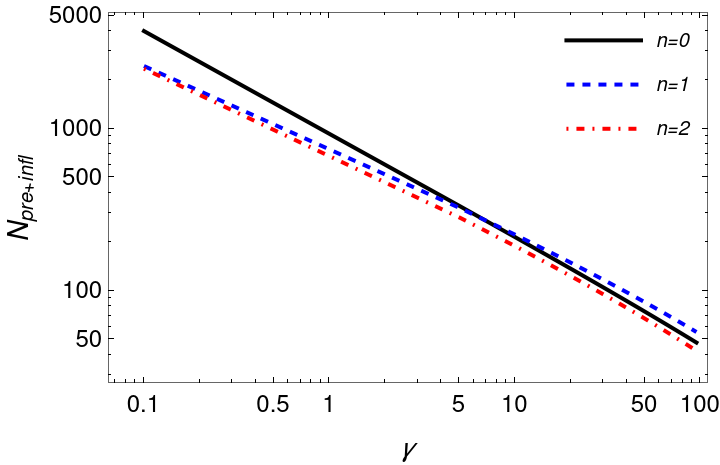} 
 \caption{Number of {\it e}-folds of evolution from the bounce up to
   the end of inflation, $N_{\rm pre+infl}$ as a function of the
   Barbero-Immirzi parameter $\gamma$. The parameter $\alpha$ in the
   potential is kept fixed at the values saturating the bound given by
   Eq.~(\ref{alphabound}) for each model.}
 \label{fig3}
 \end{figure}
\end{center}

{}Finally, we now consider the effect of varying the Barbero-Immirzi
parameter $\gamma$. As seen from the results of
{}Fig.~\ref{fig2}, small values of $\alpha$ will always lead to a larger
number of {\it e}-folds.  Since the number of {\it e}-folds of inflation has
necessarily a lower bound set by the requirement of inflation to solve
the usual flatness and horizon problems of the hot big bang model, we
fix $\alpha$ in each model such as to saturate the upper bound given
by Eq.~(\ref{alphabound}). The corresponding results are given in
{}Fig.~\ref{fig3}.

Note that in {}Fig.~\ref{fig3} we show the total number of {\it e}-folds from
the bounce up to the end of inflation, $N_{\rm pre+infl}$, i.e., we also
consider the duration of the preinflationary postbounce phase. Here,
we can take advantage  of the fact that the linear perturbations in
LQC are known analytically~\cite{Zhu:2017jew}.  In particular, from
the knowledge of the perturbation spectra in LQC, it has been shown
in Ref.~\cite{Barboza:2022hng} that there is an upper bound for the
Barbero-Immirzi parameter determined by the condition on $N_{\rm
  pre+infl}$,
\begin{equation}\label{Nbound}
N_{\rm pre+infl} \gtrsim 79 - \frac{3}{2} \ln (\gamma).   
\end{equation}
This allows us to find that the consistency of the perturbation spectra in LQC
for the $\alpha$-attractor models considered here with the
observations, 
when fixing the parameter $\alpha$ in the
   potential at the values saturating the bound given by
   Eq.~(\ref{alphabound}) for each model,
then leads to the following upper bounds on $\gamma$:
\begin{eqnarray}
\gamma \lesssim \left\{
\begin{array}{ll}
51.2, & {\rm for}\; n=0\; {\rm and} \;\alpha =10,\\ 63.4, & {\rm for}\; 
n=1\; {\rm and} \;\alpha =17,\\ 64.2, & {\rm
  for}\; n=2\; {\rm and}\; \alpha =67,
\end{array}
\right.
\label{gammabound}
\end{eqnarray}
such that larger values of $\gamma$ than these bounds would violate Eq.~(\ref{Nbound}).

\begin{center}
\begin{figure}[!htpb]
\includegraphics[width=8.2cm]{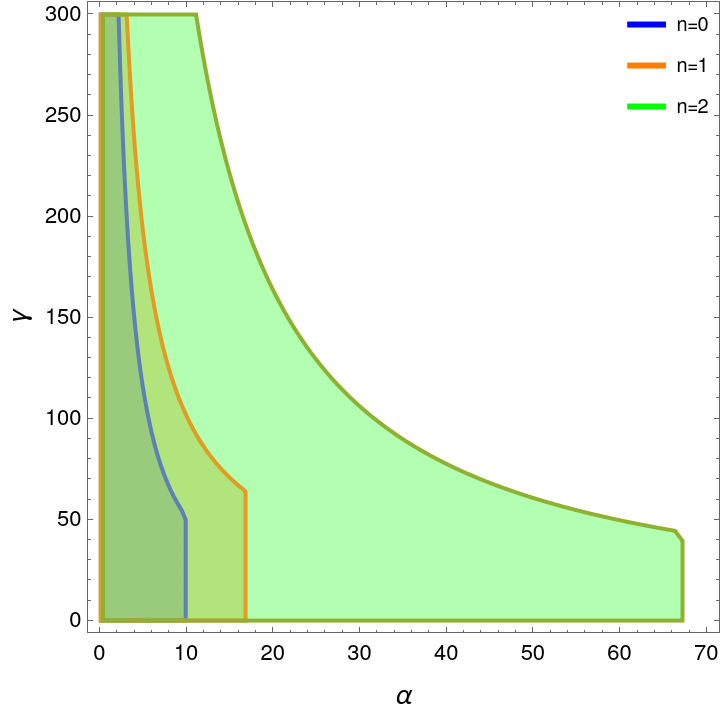} 
 \caption{The allowed region in the $\alpha$ and $\gamma$ parameter plane for
the three cases of $\alpha$-attractor potential studied here.}
 \label{alpha-gamma}
 \end{figure}
\end{center}

{}For completeness, in {}Fig.~\ref{alpha-gamma} we show the region of parameters $\alpha$
and $\gamma$ consistent with both Eq.~(\ref{Nbound}) and on the allowed range
for the tensor-to-scalar ratio and spectral tilt (see the Appendix)
for the three cases of $\alpha$-attractor potential studied here.
Note that the larger the value of $n$ is in the $\alpha$-attractor potential, 
the allowed region extends more
and more toward the right in {}Fig.~\ref{alpha-gamma}.
We can also understand the allowed region shown in {}Fig.~\ref{alpha-gamma}
by noticing that for values of $\gamma$ smaller than the upper bounds given
by Eq.~(\ref{gammabound}), the number of {\it e}-folds is sufficiently large to satisfy
the condition set by Eq.~(\ref{Nbound}). In this case, the upper bound in the
tensor-to-scalar ratio, Eq.~(\ref{valuer}), is dominant and independent of $\gamma$. 
But as $\gamma$ increases,
then the number of {\it e}-folds tends to decrease according to Eq.~(\ref{Nbound}).
Then, the $\alpha$ parameter has to be changed to compensate, leading to
a flatter potential (allowing for a larger number of {\it e}-folds).

Before closing this section, it is worth commenting that the result given by
Eq.~(\ref{Nbound}) was derived in the context of the dressed metric approach for the
perturbations in LQC~\cite{Agullo:2012sh}. Equation~(\ref{Nbound}) is obtained
using that in the perturbations derived in the dressed metric approach the spectrum
receives corrections that depend on a characteristic scale $k_B$ at the bounce,
which is the shortest scale (or more precisely, the largest wave number
$k_{B}$ that feels the spacetime curvature during the
bounce). The correction in the power spectrum can be seen as a modification
of the Bunch-Davis vacuum for the quantum fluctuations due to mode excitations 
as a consequence of the quantum bounce in LQC.
The modification of the power spectrum is constrained by the observations, which then
put a constraint on $k_B$~\cite{Zhu:2017jew,Benetti:2019kgw},
$k_{B} < 1.9 \times 10^{-4} \ {\rm Mpc}^{-1}$ at $1 \sigma$.
Since $k_{B}$ and the total number of {\it e}-folds $N_T$ from  the bounce until
today are related,
\begin{equation}\label{kbN}
k_{\rm B}\equiv \frac{\sqrt{8\pi \rho_{\rm cr}} a_{\rm B}}{m_{\rm
    Pl}}= m_{\rm Pl}\left(\frac{\sqrt{3}}{4 \pi
  \gamma^{3}}\right)^{1/2} e^{-N_{\rm T}},
\end{equation}
we can then derive the condition on $N_{\rm pre+infl}$ as given by
Eq.~(\ref{Nbound}) and detailed in Ref.~\cite{Barboza:2022hng}.

Even though Eq.~(\ref{Nbound}) was derived in the context of the
dressed metric quantization approach, the result is also qualitatively
similar when derived in the hybrid quantization approach.
In the hybrid quantization approach~\cite{Fernandez-Mendez:2012poe},
there is a similar enhancement of the power spectrum like in the
dressed case. The characteristic scale is now $k_H = k_B/\sqrt{3}$,
and the constraint from the observations now leads to~\cite{Li:2020mfi}
$k_{H} < 4.1 \times 10^{-4} \ {\rm Mpc}^{-1}$.
Overall, in this case the resulting upper bound in the number of {\it e}-folds from
the bounce until the end of inflation turns out to be similar to
Eq.~(\ref{Nbound}). Thus, our results can as well be applied to the
case of the hybrid quantization approach.\footnote{
It is also worth mentioning that the bounds on either $k_B$ or $k_H$ are a result of the deviations 
from the standard $\Lambda CDM$ model due to the quantum bounce and that 
appear at infrared scales in the CMB. These deviations of the power
spectrum with respect to the standard model and which depend on the quantization
approach used in LQC, have been studied
in several
papers~\cite{Ashtekar:2016pqn,deBlas:2016puz,Martin-Benito:2021szh,Agullo:2018wbf,Garay:2024afl}.}

\section{Conclusions}
\label{conclusions}

In this paper, we have revisited some of the results concerning a
class of $\alpha$-attractor potentials in the context of LQC. We have
considered the background evolution for these types of potentials by
following the dynamics when setting the initial conditions at the
bounce and also in the deep contracting phase before the bounce. The
latter case has been claimed in the recent literature to be the
correct point where the initial conditions should be set. It has been
shown in several
references~\cite{Linsefors:2013cd,Linsefors:2014tna,Bolliet:2017czc,Martineau:2017sti,Barboza:2020jux}
that in this case the inflationary evolution can be predicted.  By taking
advantage of the results obtained in Ref.~\cite{Barboza:2022hng}, we
have studied the dynamics of the $\alpha$-attractor models when
varying the potential parameter $\alpha$ and also  the Barbero-Immirzi
parameter $\gamma$.  By contrasting the results with the observations,
we were able to put constraints on both of these parameters. In
particular, the tensor-to-scalar ratio imposes the upper bounds in the
$\alpha$ parameter as being, $\alpha_{n=0} \lesssim 10$, $\alpha_{n=1}
\lesssim 17$, and $\alpha_{n=2} \lesssim 67$,  for the
$\alpha$-attractors T, E, and $n=2$ models, respectively.

Our results have also shown that by decreasing the $\alpha$ parameter,
the number of {\it e}-folds predicted for each model in LQC
increases. Likewise, the lower the value for the Barbero-Immirzi parameter is,
the larger also is the number of {\it e}-folds of inflation. Then, by a
previous general bound determined in Ref.~\cite{Barboza:2022hng} and
set on the total duration of the preinflationary postbounce  and
later inflationary phases in LQC, we were able to find the further
upper bounds for the Barbero-Immirzi parameter for the
$\alpha$-attractor models,
when fixing the parameter $\alpha$ in the
   potential at the values saturating the upper bounds given above for each model: 
$\gamma_{n=0} \lesssim 51.2$, $\gamma_{n=1}
\lesssim 63.4$, and $\gamma_{n=2} \lesssim 64.2$.  Our results show 
that when we combine constraints on the observables and
predictions about the duration of inflation in LQC, general bounds can
be set on the $\alpha$-attractor potentials and also on the
Barbero-Immirzi parameter.  

In general, for most of the parameter region for the allowed values
for $\alpha$ and for the Barbero-Immirzi
parameter, our results predict a large number of {\it e}-folds for the 
$\alpha$-attractor potentials studied here. 
A large number of {\it e}-folds is typically associated with the so-called 
trans-Planckian problem~\cite{Martin:2000xs,Brandenberger:2012aj}
(for a more recent discussion, made in connection with the swampland
conjectures of quantum gravity, see, e.g.,~\cite{Bedroya:2019snp,Bedroya:2019tba}).
Inflation models with a large number of {\it e}-folds are also associated with
the presence of eternal inflation~\cite{Guth:2000ka}, which can be considered typical
for potentials with very flat plateaus, like the ones considered here.
Possible observational consequences of the trans-Planckian  problem in the
context of LQC have also been recently considered in Ref.~\cite{Garay:2024afl},
indicating that the results are very much model dependent. 
Independent of the formal issues related to the trans-Planckian  problem,
we recall that there are some models that in fact require a large number of {\it e}-folds,
like in the relaxion inflation model~\cite{Graham:2015cka}, in stochastic 
axion scenarios~\cite{Graham:2018jyp,Ho:2019ayl,Takahashi:2019pqf}, and in some
quintessence models~\cite{Ringeval:2010hf,Ringeval:2019bob}. 
We believe that delving into the issues that the prediction of a large number of
{\it e}-folds in the models studied here might present is beyond the
scope of our work, but this is certainly something worth exploring in the future.

\section*{Acknowledgements}

G.L.L.W.L. acknowledges financial support of the 
Coordena\c{c}\~ao de Aperfei\c{c}oamento de Pessoal de N\'{\i}vel
Superior (CAPES)  - Finance code 001.    R.O.R. acknowledges financial
support by research grants from Conselho Nacional de Desenvolvimento
Cient\'{\i}fico e Tecnol\'ogico (CNPq), Grant No. 307286/2021-5, and
from Funda\c{c}\~ao Carlos Chagas Filho de Amparo \`a Pesquisa do
Estado do Rio de Janeiro (FAPERJ), Grant No. E-26/201.150/2021.   

\section*{Appendix}
\label{AppA} 

The normalization $V_0$ of the inflaton potential is fixed by the
amplitude of the CMB primordial scalar of curvature power spectrum
$\Delta_{{\cal R}}$, given by~\cite{Lyth:2009zz}
\begin{eqnarray} \label{PkCI}
\Delta_{{\cal R}} &=&  \left(\frac{ H_*^2}{2
  \pi\dot{\phi}_*}\right)^2,
\end{eqnarray}
where a subindex $*$ means that the quantities are evaluated at the
Hubble radius crossing  $k_*$  ($k_*=a_* H_*$), which typically
happens around $N_* \sim 50-60$ {\it e}-folds before the end of
inflation.  {}From the Planck Collaboration~\cite{Aghanim:2018eyx},
$\ln\left(10^{10} \Delta_{{\cal R}} \right) \simeq 3.047$
(TT,TE,EE-lowE+lensing+BAO 68$\%$ limits).  During the slow-roll
regime of inflation, we have $H^{2} \simeq 8\,\pi V/( 3\,m_{\rm
  Pl}^{2})$  and $\dot{\phi} \simeq -V_{,\phi}/(3\,H)$, which then
gives for Eq.~(\ref{PkCI}) the result
\begin{equation}\label{spectrum_amplitude}
 \Delta_{\rm R} \simeq \frac{128\,\pi}{3\,m_{\rm Pl}^{6}}
 \frac{V_{\ast}^{3}}{V_{,\phi \ast}^{2}} . 
\end{equation}

By fixing $N_*$, we can determine $\phi_*$ by solving the number of
{\it e}-folds equation,
\begin{equation}
   N_* = \frac{8\pi}{m_{\rm Pl}^2}  \int_{\phi_{\rm end}}^{\phi_*}
   \frac{V}{V'} d\phi.
\label{N*}
\end{equation}
{}For the $\alpha$-attractor class of potentials considered here,
Eq.~(\ref{alphattractor}), $\phi_{\rm end}$ is given by
Eq.~(\ref{phiend}) and $N_*$ is found to be given by
%
\begin{eqnarray}
N_* &=& \frac{3 \alpha}{4 n(n-2)^2 } \left\{2 b (n-2)^2-(n-2) \left[n \left(e^{2
      b}-y_* \right)
\right. \right.
\nonumber \\
&+& \left. \left. (n-2) \ln(y_*)\right]-4 (n-1) \left(
  \ln\left[e^{2 b} (n-2)+n\right]
\right.\right.
\nonumber \\
&-& \left. \left. \ln(n+2y_*-n y_*)\right)\right\}, \nonumber \\
\label{Nstar}
\end{eqnarray}
%
where we have defined
\begin{equation}
y_*= e^{\frac{2}{3 \alpha} \frac{\phi_*}{m_{\rm Pl}}},
\end{equation}
\begin{equation}
b= {\rm arccoth}(a_n),
\end{equation}
and
\begin{equation}
a_n= \frac{1}{2} \left[n+\sqrt{3 \alpha} +\sqrt{(n-2)^2+ 2
    n\sqrt{3\alpha}+3 \alpha }\right].
\end{equation}

\begin{center}
\begin{figure}[!htpb]
\includegraphics[width=8.2cm]{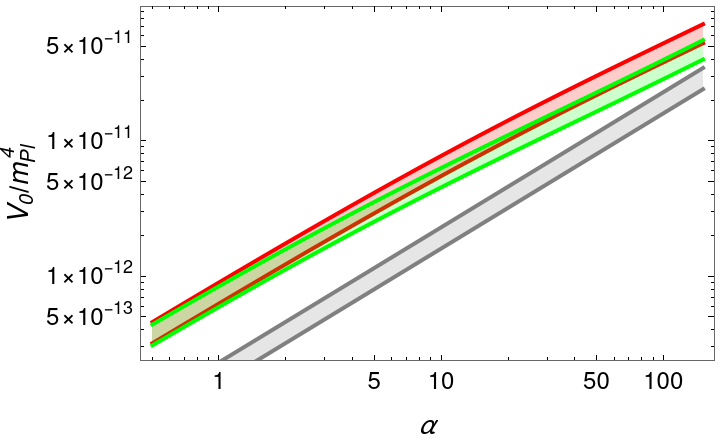}
 \caption{The normalization $V_0$ for the $\alpha$-attractor
   potential models considered in this work.  The gray, green, and red
   curves and respective regions are for the T-model ($n=0$), E-model ($n=1$) and
   $n=2$ $\alpha$-attractor models, respectively. They are obtained by fixing
   $N_*$ in the values $N_*=50$ (upper curve) and $N_*=60$ (lower
   curve).}
 \label{fig:V0alpha}
 \end{figure}
\end{center}

{}From Eq.~(\ref{Nstar}), we then find that for the T model ($n=0$)
\begin{eqnarray}
\phi_*^{n=0} = \frac{m_{\text{Pl}}}{4} \sqrt{\frac{3\alpha}{\pi }}
\text{ArcCosh}\left[ \frac{4 N_*}{3 \alpha } -
  \frac{1+a_0^2}{1-a_0^2}\right] ,
\label{phi*n0}
\end{eqnarray}
for the E model ($n=1$), we have that
%
\begin{eqnarray}
\phi_*^{n=1} &=& \sqrt{\frac{3 \alpha}{\pi}} \frac{m_{\text{Pl}}}{4}
\left\{\frac{1 +2  \text{arccoth} \left(a_1\right)}{a_1-1} \right.
\nonumber\\ &-&\left.  W_{-1}\left[\frac{e^{-1+\frac{2}{a_1-1}-
      \frac{4N_*}{3 \alpha }} \left(1+a_1\right)}{a_1-1}\right]
\right.  \nonumber \\ &-& \left.\frac{4 N_*}{3(a_1-1) \alpha} \right.
\nonumber \\ &+&\left. \frac{a_1}{a_1-1} \left(1 -2
\text{arccoth}\left(a_1\right)+\frac{4 N_*}{3 \alpha}\right)\right\},
\label{phi*n1}
\end{eqnarray}
%
and for $n=2$, we have that
%
\begin{eqnarray}
\!\!\!\!\!\!\!\!\phi_*^{n=2} &=& \frac{m_{\text{Pl}}}{24 \sqrt{\pi
    \alpha }} \left\{-12 \sqrt{\alpha } -16 \sqrt{3} N_* \right.
\nonumber \\ &+& \left. 3 \sqrt{3} \alpha   \left(-1+4
\text{arccoth}\left(a_2\right) \right. \right.  \nonumber \\ &-&
\left.\left.  W_{-1}\left[-e^{-1-\frac{4}{\sqrt{3} \sqrt{\alpha }}+4
    \text{arccoth}\left(a_2\right)-\frac{16 N_*}{3 \alpha }}
  \right]\right)\right\},
\label{phi*n2}
\end{eqnarray}
%
where in the above equations, $W_{-1}(x)$ is the Lambert function.

\begin{center}
\begin{figure*}[!htpb]
\subfigure[]{\includegraphics[width=8.2cm]{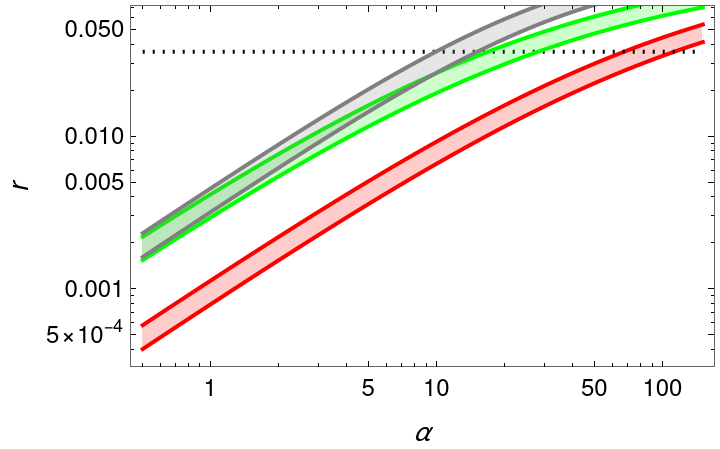}}
\subfigure[]{\includegraphics[width=8.2cm]{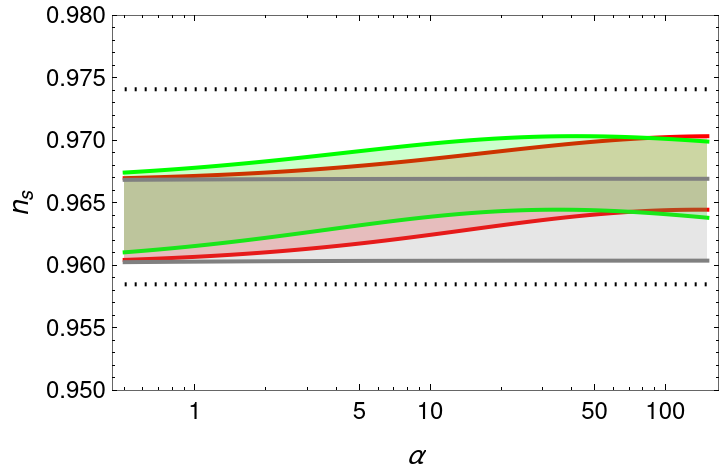} }
 \caption{The tensor-to-scalar ratio (a) and the spectral tilt
   (b)  for the $\alpha$-attractor potential models considered
   in this work.  The gray, green, and red curves and respective regions are for
   the T-model ($n=0$), E-model ($n=1$), and $n=2$ $\alpha$-attractor
   models, respectively. The upper curves correspond to $N_*=50$,
   while the lower curves are for $N_*=60$. The horizontal dotted
   lines mark the upper bound for $r$ [given by Eq.~(\ref{valuer})] and the
   upper and lower ranges at 2-$\sigma$ for $n_s$.}
 \label{fig:rns}
 \end{figure*}
\end{center}

Note that the above equations depend on the value of $\alpha$.  In
{}Fig.~\ref{fig:V0alpha} we show the potential normalization $V_0$ as
a function of $\alpha$ when making use of
Eq.~(\ref{spectrum_amplitude}).

The use of the tensor-to-scalar ratio $r$ and the spectral tilt $n_s$
of the scalar spectrum allows us to find an upper bound for the
parameter $\alpha$ for each of the potentials considered here.  $r$
and $n_s$ are defined by~\cite{Lyth:2009zz}
\begin{equation}
r=16 \epsilon_V,
\end{equation}
and
\begin{equation}
n_s= 1-6 \epsilon_V + 2 \eta_V,
\end{equation}
where $\epsilon_V$ and $\eta_V$ are the slow-roll coefficients:
\begin{equation}
\epsilon _V=\frac{m_{\rm Pl}^2}{16 \pi }
\left(\frac{V'}{V}\right)^{2},\;\;\;\; \eta_V= 
\frac{m_{\rm Pl}^2}{8 \pi } \frac{V''}{V},
\end{equation}
and which are evaluated at the value $\phi_*$.

{}From the expressions (\ref{phi*n0}) -- (\ref{phi*n2}), we evaluate
both $r$ and $n_s$ for the fiducial values of $N_*=50$ and $N_*=60$ as
a function of $\alpha$. The results are shown in {}Fig.~\ref{fig:rns}.
{}For $\alpha \ll 1$, the results approach the ones given in
Eqs.~(\ref{r}) and (\ref{ns}), for $r$ and $n_s$, respectively.  {}For
$\alpha > 1$, the tensor-to-scalar ratio gives the stronger constrain
on the value for $\alpha$ for each type of potential, leading to the
values given in Eq.~(\ref{alphabound}). 


\end{document}